\documentclass[11pt]{article}
\usepackage{latexsym}
\usepackage{amssymb}
\usepackage{epsf}
\usepackage{amsmath}
\usepackage[hypertex]{hyperref}
\usepackage{graphicx}

\newcommand{\tr}{{\rm Tr}}

\begin{document}

\pagestyle{empty}

\begin{flushright}
SLAC-PUB-12252\\
CALT-68-2621 \\
hep-ph/0612139 \\
\end{flushright}

\vspace{2.5cm}

\begin{center}

{\bf\Large Direct Mediation of Meta-Stable Supersymmetry Breaking}
\\

\vspace*{1.5cm}
{\large 
Ryuichiro Kitano$^{1,2}$,
Hirosi Ooguri$^3$ and
Yutaka Ookouchi$^3$} \\
\vspace*{0.5cm}

$^1${\it Stanford Linear Accelerator Center, Stanford University,
                Stanford, CA 94309} \\
$^2${\it Physics Department, Stanford University, Stanford, CA 94305}\\
$^3${\it California Institute of Technology, Pasadena, CA 91125}
\vspace*{0.5cm}

\end{center}

\vspace*{1.0cm}

\begin{abstract}
{
\baselineskip 14pt

The supersymmetric SU($N_C$) Yang-Mills theory coupled to $N_F$ matter
fields in the fundamental representation has meta-stable vacua with
broken supersymmetry when $N_C < N_F < {3\over 2} N_C$. By gauging the
flavor symmetry, this model can be coupled directly to the standard
model. We show that it is possible to make a slight deformation to the
model so that gaugino masses are generated and the Landau pole problem
can be avoided.  The deformed model has simple realizations on
intersecting branes in string theory, where various features of the
meta-stable vacua are encoded geometrically as brane configurations.

}
\end{abstract} 

\newpage
\baselineskip 18pt
\setcounter{page}{2}
\pagestyle{plain}
\section{Introduction}

Although there is no clear evidence yet, it is plausible that softly
broken ${\cal N}=1$ supersymmetry is realized in nature. Not only because it is
a symmetry possessed by string theory, there are many phenomenologically
attractive features in supersymmetric models, such as cancellation of
quadratic divergences and unification of the gauge coupling
constants~\cite{Dimopoulos:1981zb,Dimopoulos:1981yj,Sakai:1981gr}.

It is then a question how supersymmetry is broken and how we feel
it. There have been many studies on this subject, but, as is often the
case, one of the earliest proposals~\cite{Dine:1981za,Dimopoulos:1981au}
among them seems to be the most elegant and simple idea. The idea is
that there is a QCD-like strong interaction which breaks supersymmetry
dynamically, and the standard model gauge group is identified with a
subgroup of flavor symmetry in this sector. The standard model gauge
sector can, therefore, feel the supersymmetry breaking directly via
one-loop diagrams.

This idea has been discarded for a long time because of its difficulty
in realistic model building.
First, Witten has shown that there is a supersymmetric vacuum in
supersymmetric QCD by using an index
argument~\cite{Witten:1982df}. Therefore, we are forced to think of the
possibility of chiral gauge theories for supersymmetry breaking, which
is already a bit complicated. (See
\cite{Affleck:1983rr,Affleck:1983mk,Affleck:1984xz} for dynamical
supersymmetry breaking in chiral gauge theories, and
\cite{Poppitz:1996fw,Arkani-Hamed:1997jv} for models of direct gauge
mediation in that context.)
There is also a problem of Landau poles of the standard model gauge
interactions. Once we embed the gauge group of the standard model into a
flavor group of the dynamical sector (this itself is not a trivial
task), there appear many particles which transform under the standard
model gauge group. These fields contribute to beta functions of the
gauge coupling constants and drive them to a Landau pole below the
unification scale.
Finally, even though the gauge sector of the standard model directly
couples to the supersymmetry breaking dynamics, it is non-trivial
whether we can obtain the gaugino masses. It is often the case that
the leading contribution to gaugino masses cancels out.

Very recently, there was a break-through on this subject. Intriligator,
Seiberg and Shih (ISS) have shown that there {\it is} a meta-stable
supersymmetry breaking vacuum in some of supersymmetric QCD
theories~\cite{Intriligator:2006dd}. The model is simply SU($N_C$) gauge
theory with massive (but light) $N_F$ quarks. Within a range $N_C < N_F
< {3\over 2}N_C$, supersymmetry is broken in the meta-stable vacuum.
The possibility of direct gauge mediation in this model is also
discussed in Ref.~\cite{Intriligator:2006dd}.\footnote{See also
\cite{Banks:2006ma} for a related work.} Because of its simplicity of
the model, it is straightforward to embed the standard model gauge group
into the SU($N_F$) flavor symmetry. However, it is concluded that there
are still problems regarding the Landau pole and the gaugino masses.
In the ISS model, there is an unbroken approximate U(1)$_R$ symmetry
which prevents us from obtaining the gaugino masses.

The U(1)$_R$ problem is a common feature in models of gauge
mediation. As is discussed recently in Ref.~\cite{Dine:2006xt}, if the
low energy effective theory of the dynamical supersymmetry breaking
model is of the O'Raifeartaigh type, there is an unbroken $R$-symmetry
at the minimum of the potential (the origin of the field space).
It has been proposed that the inverted hierarchy
mechanism~\cite{Witten:1981kv} can shift the minimum away from the
origin by the effect of gauge
interactions~\cite{Murayama:1997pb,Dimopoulos:1997je,Luty:1997ny,Agashe:1998wm,Dine:2006xt}.
An alternative possibility that the shift is induced by an $R$-symmetry
breaking term in supergravity Lagrangian (the constant term in the
superpotential) is recently discussed in Ref.~\cite{Kitano:2006wz}.
It is, however, still non-trivial whether we obtain the gaugino masses
even with the $R$-symmetry breaking vacuum expectation values in direct gauge mediation
models. For example, a model in Ref.~\cite{Izawa:1997gs} generates
gaugino masses only at the $F^3$ order even though the $R$-symmetry is
broken by assuming the presence of the local minimum away from the
origin. Since the scalar masses squared are obtained at the $F^2$ order
as usual, gaugino masses are much smaller than the scalar masses unless
the messenger scale is $O(10~{\rm TeV})$, that is difficult in models of
direct gauge mediation because of the Landau pole problem.
In fact, as we will see later, the structure of the messenger particles
in the ISS model is the same as that in this model. (The same structure
can be found in many models, for example, in Ref.~\cite{Kitano:2006wm}
and also in very early proposals of gauge mediation models in
Ref.~\cite{Dine:1981gu,Nappi:1982hm,Alvarez-Gaume:1981wy}.)
Therefore, it is not sufficient to destabilize the origin of the field
space for generating both gaugino and scalar masses.

In this paper we propose a slight deformation to the ISS model with
which we can obtain gaugino masses by identifying a flavor subgroup with
the standard model gauge group.
We add a superpotential term which breaks $R$-symmetry explicitly so
that non-vanishing gaugino masses are induced. The vacuum structure
becomes richer by the presence of the new term. In addition to the
vacuum that is obtained by a slight perturbation to the ISS meta-stable
vacuum, which we will call the ISS vacuum, there appear 
new (but phenomenologically unacceptable) meta-stable vacua.
We find that decays of the ISS vacuum into the other vacua are
sufficiently slow so that it is phenomenologically viable.

We also show that the Landau pole problem can be avoided by keeping the
dynamical scale of the ISS sector sufficiently high in a way that is
compatible with phenomenological requirements. In addition, if
meta-stable vacua exist in a model with the same number of colors and
flavors, as suggested by ISS, we can also consider the case where the
ISS sector is in the conformal window, ${3 \over 2} N_C \leq N_F < 3
N_C$. In this case, we can take the scales of the ISS sector as low as
$O$(100--1000 TeV).

The deformed ISS model can be realized on intersecting branes in string
theory, where a rich vacuum structure and the meta-stability of vacua
can be understood geometrically.

\section{The ISS model}

We first review the ISS model. The model is simply a supersymmetric QCD
with light flavors. Perturbative corrections to a scalar potential are
calculable in the magnetic dual picture, and they have been
found to stabilize a
supersymmetry breaking vacuum. The model has an unbroken $R$-symmetry, which
prevents it from generating gaugino masses. An explicit one-loop computation
of the masses suggests a natural solution to this problem, which we will 
discuss in the next section. 

\subsection{Supersymmetry breaking}

The model is an SU($N_C$) gauge theory with $N_F$ flavors. The quarks
have mass terms:
\begin{eqnarray}
 W = m_i Q_i \bar Q_i \ .
\end{eqnarray}
The index $i$ runs for $i = 1, \cdots, N_F$. The masses $m_i$ are assumed to be
much smaller than the dynamical scale $\Lambda$.
There is a meta-stable supersymmetry breaking vacuum when $N_C < N_F <
{3\over 2}N_C$, where there is a weakly coupled description of the theory
below the dynamical scale $\Lambda$.
The gauge group of the theory is SU($N_F - N_C$) and degrees of freedom
at low energy are meson fields $M_{ij} \sim Q_i Q_j$ and dual quarks
$q_i$ and $\bar q_i$.  There are superpotential terms:
\begin{eqnarray}
 W = m_i M_{ii} - \frac{1}{\hat \Lambda} q_i M_{ij} \bar q_j\ .
\end{eqnarray}
A dimensionful parameter $\hat \Lambda$ is introduced so that the
dimensionality of the superpotential is correct. A natural scale of
$\hat \Lambda$ is $O(\Lambda)$.

With this superpotential, the $F_M = 0$ condition for all components of
 $M_{ij}$ cannot be satisfied. The rank of the matrix $q_i \bar q_j$ is
 at most $N_F - N_C$ whereas the mass matrix $m_{i}$ has the maximum
 rank, $N_F$.
The lowest energy vacuum is at 
\begin{eqnarray}
 M_{ij} = 0\ ,\ \ \ 
q_i = \bar q_i = \left(
\begin{array}{c}
 \sqrt{m_I \hat \Lambda} \ \delta_{IJ} \\
 0
\end{array}
\right)\ ,
\end{eqnarray}
where $I$ and $J$ runs from 1 to $N_F - N_C$, and $m_i$ is sorted in
descending order. The $F$-components of $M_{ii}$ with $i = N_F-N_C+1,
\cdots, N_F$ have non-vanishing value $m_i$.
At this vacuum, the gauge symmetry SU($N_F - N_C$) is completely broken.

We parametrize fluctuations around this vacuum to be:
\begin{eqnarray}
\frac{ \delta M_{ij} }{ \hat \Lambda } = h
\left(
\begin{array}{cc}
 Y_{IJ} & Z_{I a} \\
 \tilde Z_{aI} & \hat \Phi_{ab} \\
\end{array}
\right)\ ,\ \ \ 
\delta q_i =
\left(
\begin{array}{c}
 \chi_{IJ} \\
 \rho_{Ia} \\
\end{array}
\right)
\ , \ \ \ 
\delta \bar q_i =
\left(
\begin{array}{c}
 \tilde \chi_{IJ} \\
 \tilde \rho_{Ia} \\
\end{array}
\right)\ .
\end{eqnarray}
We put dimensionless parameter $h$ of $O(1)$ so that components have
canonically normalized kinetic term.
Again, $I,J = 1, \cdots, N_F - N_C$ and $a,b = 1, \cdots, N_C$.
Among these fields $\hat \Phi_{ab}$ and the trace part of $\chi -
\tilde \chi$, $\tr [\chi - \tilde \chi] \equiv \tr \delta \hat \chi$,
remains massless at tree level. The other fields obtain masses of
$O(\sqrt{m \Lambda})$.
One-loop correction to a potential for the pseudo-moduli $\hat \Phi$ and
Re$[\tr \delta \hat \chi]$ is shown to give positive masses squared,
which ensures the stability of the vacuum.\footnote{Imaginary part of
$\tr \delta \hat \chi$ is a Goldstone boson associated with a broken
U(1)$_B$ symmetry.}
Once we take into account the non-perturbative effect, the true
supersymmetric vacuum appears far away from the origin of the meson
field $M$. The life-time of the false vacuum can be arbitrarily long if
$m_i \ll \Lambda$. Also, interestingly, the supersymmetry breaking
vacuum is preferred in the thermal history of the universe~
\cite{reheatingone,reheatingtwo,reheatingthree}.

\subsection{Gaugino masses}

It is possible to embed the standard model gauge group into a flavor
symmetry group of this model.
When we take $m_1 = \cdots = m_{N_F - N_C} = m $ and $m_{N_F - N_C +1} =
\cdots = m_{N_F} = \mu$, there is a global symmetry; SU($N_F - N_C$)$_F$
$\times$ SU($N_C$)$_F$ $\times$ U(1)$_B$. With $N_F - N_C \geq 5$ or
$N_C \geq 5$, we can embed SU(3) $\times$ SU(2) $\times$ U(1) into
SU($N_F - N_C$)$_F$ or SU($N_C$)$_F$, respectively.
In the case where we embed SU(3) $\times$ SU(2) $\times$ U(1) into the
SU($N_F - N_C$)$_F$ flavor symmetry, the standard model gauge group at low
energy is a diagonal subgroup of SU(3) $\times$ SU(2) $\times$ U(1) in
SU($N_F - N_C$) dual gauge interaction (under which $q$ and $\bar q$
transform and $M$ is neutral) and that in the SU($N_F - N_C$)$_F$ flavor
group.

As discussed in Ref.~\cite{Intriligator:2006dd}, there is an unbroken
$R$-symmetry under which $M$ carries charge two and $q$ and $\bar q$ are
neutral.
Since the $R$-symmetry forbids the gaugino masses, there is no
contribution to the gaugino masses of the standard model gauge group
even though it is directly coupled to a supersymmetry breaking sector.
It is instructive to see how the gaugino masses vanish at one-loop.  The
fields $\rho$ and $\tilde \rho$ carry quantum numbers of both SU($N_F -
N_C$) and SU($N_C$)$_F$ and couple to $\hat{\Phi}$ which has
non-vanishing vacuum expectation value in the $F$-component.
Therefore $\rho$ and $\tilde \rho$ play a role of messenger fields in
gauge mediation.\footnote{The standard model gauge group at low energy
partly comes from SU($N_F - N_C$) when we embed the SU(3) $\times$ SU(2)
$\times$ U(1) into SU($N_F - N_C$)$_F$. One-loop diagrams with the
$\rho$ and $\tilde \rho$ fields, therefore, contribute to the gaugino
masses also in this case, although they are not charged under SU($N_F-
N_C$)$_F$.}
The relevant superpotential for this discussion is 
\begin{eqnarray}
 W = - h \rho \hat \Phi \tilde \rho
- h \bar m ( \rho \tilde Z + \tilde \rho Z )\ ,
\end{eqnarray}
where we suppressed indices and defined $\bar m \equiv \sqrt{m \hat
\Lambda}$.
The $\rho$ and $Z$ fields have mixing terms. In a matrix notation,
\begin{eqnarray}
 W = h ( \rho , Z) {\cal M}
\left(
\begin{array}{c}
 \tilde \rho \\
 \tilde Z\\
\end{array}
\right)
\end{eqnarray}
where ${\cal M}$ is a mass matrix for the messenger fields
\begin{eqnarray}
{\cal M} = \left(
\begin{array}{cc}
 \hat \Phi & \bar m \\
 \bar m & 0 \\
\end{array}
\right)\ .
\end{eqnarray}
The formula for the gaugino masses can be generalized for this
multi-messenger case as follows:
\begin{eqnarray}
 m_\lambda = \frac{g^2 \bar N}{(4 \pi)^2} 
F_{\hat \Phi} \frac{\partial}{\partial \hat \Phi} \log \det {\cal M}\ ,
\label{eq:gaugino}
\end{eqnarray}
where $\bar N$ is $N_C$ or $N_F - N_C$ depending on whether we embed the
standard model gauge group into the SU($N_F - N_C$)$_F$ or the
SU($N_C$)$_F$ flavor symmetry.
This formula is valid when $F_{\hat \Phi} \ll \bar m^2$.
Since there is no $\hat \Phi$ dependence in $\det {\cal M}$, we obtain
$m_\lambda = 0$.

We can now clearly see that the gaugino mass would vanish at the leading
order in $F_\Phi /\bar m^2$ even if we could obtain a non-vanishing vacuum 
expectation value for $\hat
\Phi$ which breaks the $R$-symmetry~\cite{Izawa:1997gs}. In the
following section, we consider a model with explicit $R$-symmetry
breaking which generates the gaugino masses at the leading order in $F_\Phi /
\bar m^2$.

\section{Deformed ISS model}

Motivated by discussion in the previous section, we consider a
modification of the ISS model which contains a mass term for the meson
fields $Z$ and $\tilde Z$ so that $\det {\cal M}$ has $\hat \Phi$
dependence.
In the electric description, this corresponds to adding the following
superpotential term
\begin{eqnarray}
 W \ni - \frac{1}{m_X} (Q_I \bar Q_a) (Q_a \bar Q_I)\ ,
\label{eq:higher-d}
\end{eqnarray}
where the color SU($N_C$) indices are contracted in $(Q \bar Q)$. 
Though this is a non-renormalizable interaction, it can be generated by
integrating out extra massive fields coupled to $(Q_a, Q_I)$ in 
a renormalizable theory. In section 4, we will show that such a theory
can be realized on intersecting branes in string theory.
This interaction preserves the global symmetry SU($N_F - N_C$)$_F$
$\times$ SU($N_C$)$_F$ $\times$ U(1)$_B$. We assume the same structure
for mass terms of $Q$ and $\bar Q$ as that in the model in the previous
section, i.e.,
\begin{eqnarray}
 W_{\rm mass} = m (Q_I \bar Q_I) + \mu (Q_a \bar Q_a)\ ,
\end{eqnarray}
so that the global symmetry is preserved. In the magnetic description, 
the mass terms correspond
\begin{eqnarray}
 W_{\rm mass}^{\rm mag.} = \bar m^2 \tr Y + \bar \mu^2 \tr \hat{\Phi}\ ,
\end{eqnarray}
where $\bar m^2 \equiv m \hat \Lambda$ and $\bar \mu^2 \equiv \mu \hat
\Lambda$.
In terms of component fields, the full superpotential is given by
\begin{eqnarray}
W=h \tr \left[ 
\bar m^2 Y
+ \bar \mu^2 \hat{\Phi}
- \chi Y \tilde{\chi}-\chi Z \tilde{\rho}-\rho \tilde{Z} \tilde{\chi}
-\rho \hat{\Phi} \tilde{\rho}  
-m_z Z \tilde{Z} \right].
\end{eqnarray}

We could have added other terms compatible with the global
symmetry. Although the theorem of \cite{NS} implies that a generic
deformation to the superpotential generates a supersymmetry preserving
vacuum at tree-level, it may not cause a problem with our scenario as
far as the new vacuum is far from the one we are interested in and the
transition rate between the vacua is small. However, since there are
tree-level flat directions in $\hat{\Phi}$, a deformation by $\tr
\hat{\Phi}^2$ destabilizes the ISS vacuum.
Whether such a deformation is prohibited is a question 
of ultra-violet completions of the theory, but there is an
interesting observation we can make from the point of view of
the low energy effective theory.
As we will see later, we need a certain level of hierarchy between $m$
and $\mu$ ($\mu \ll m$) to suppress a tunneling rate into unwanted vacua
and also to avoid a Landau pole of the gauge coupling of the standard
model gauge interaction.
With this hierarchy this model possesses an approximate (anomalous)
$R$-symmetry which is softly broken by the small mass term $\mu$.
The charge assignment is $R(Q_I) =R(\bar Q_I) =1 $ and $ R(Q_a) = R(\bar
Q_a) = 0$. This symmetry justifies the absence or suppression of other
higher dimensional operators such as $\tr \hat \Phi^2$ which 
destabilize the supersymmetry breaking vacua. (The supersymmetry breaking
vacua remain stable as far as the coefficient for $\tr \hat \Phi^2$ 
is smaller than $\mu$.)

\subsection{Vacuum structure}

The introduction of the mass term for $Z$ and $\tilde Z$ makes the
vacuum structure of this model quite rich. In addition to the
supersymmetric and supersymmetry breaking vacua in the ISS model, there
are also several stable supersymmetry breaking vacua. The stability and
decay probability between these vacua are controlled by parameters in
superpotential.

\vspace{0.3cm}
{\it \large Meta-stable supersymmetry breaking vacua}

\begin{figure}[htbp]
\begin{center}
  \epsfxsize=7cm \epsfbox{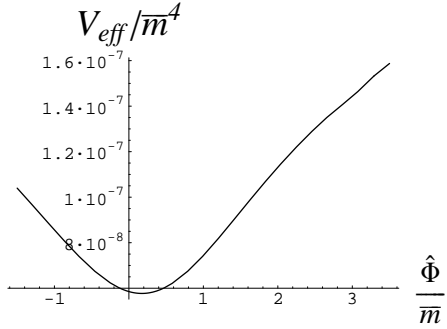}
\end{center}
 \vspace{-.5cm}
\caption{\sl One-loop effective potential $ V_{eff}(\hat \Phi/ \bar m)$
 along real axis of $\hat \Phi$ for {\rm Re}$[\tr \delta \hat \chi ] = 0$, $m_z=
 \bar m /3$ and $\bar \mu= \bar m / 100$. The critical point is at $\hat
 \Phi=0.1747 \bar m$. At $\hat \Phi \sim 3 \bar m$ there is a tachyonic
 direction toward non-zero $\rho$, $\tilde{\rho}$, $Z$ and $\tilde{Z}$
 in field configuration space.  \label{fig:potential}
}
\end{figure}

\noindent 
As long as $m_z$ is smaller than $\bar m$, we can think of $m_z$ as
a small perturbation to the ISS model, and thus there exists a similar
meta-stable supersymmetry breaking vacuum.
The effect of a finite value of $m_z$ is a small shift of $\hat \Phi$ of
$O(m_z)$.
The pseudo-moduli $\hat \Phi$ and Re$[\tr \delta \hat \chi]$ obtain masses
of $O(h^2 \bar \mu^2 / \bar{m})$ as in the ISS model.
We show in Figure~\ref{fig:potential} the one-loop effective potential
for the pseudo-moduli $\hat \Phi$. We see a small shift of the minimum.
For $m_z > \bar m$, this vacuum is destabilized. Therefore, we assume in
the following that $m_z$ is smaller than $\bar m$.

%

The small $m_z$, in fact, modifies the vacuum structure drastically at
far away from the origin of the field space. We can find other
supersymmetry breaking vacua with
\begin{align}
\label{mstable} 
\rho \tilde{\rho}
=\frac{m_z^2}{ m^2}Z\tilde{Z}= {\rm diag}(\bar \mu^2, \dots \bar \mu^2,0 \dots 0),  \quad
\chi \tilde{\chi}= \bar m  {\bf 1}_{N_F-N_C},  \quad
\end{align}
\begin{align}
\label{mstable2} 
Y = - \frac{\bar \mu^2}{m_z} {\bf 1}_{N_F - N_C}, \quad
\hat \Phi = - \frac{\bar m^2}{m_z} {\rm diag} (1, \dots 1, 0, \dots 0)\ , \quad
 V_{\rm lower}=(N_C-n)|h \bar \mu^2|^2\ ,
\end{align}
where the number of $\bar \mu^2$ in the first equation, denoted $n$,
runs from $1$ to $N_F-N_C$.  Since these vacua have energy that are
lower than that of the ISS vacua, $V_{\rm ISS}=N_C |h \bar \mu^2|^2$, 
it has non-zero transition probability to these vacua. Below, we show 
that the decay rate can be made parametrically small by a mass hierarchy, 
$\bar \mu \ll \bar m$. Although the vacuum with $n=N_F-N_C$ is the global minimum of
the classical potential, they are not phenomenologically viable since
gauginos cannot get masses at the leading order in $F/\bar m^2$ for the
same reason as in the original ISS model when we embed the standard
model into some of unbroken global symmetry.

As we have seen, our vacuum is not the global minimum of the
potential. It can decay into lower energy vacua specified by
(\ref{mstable}) and (\ref{mstable2}).
We estimate the decay rate by evaluating the Euclidean action from our
vacuum to others. The barrier by the one-loop potential is not high, of
order $O(\bar \mu^4)$. Thus, the most efficient path is to climb up the
potential of $\hat \Phi$ and then slide down to more stable
supersymmetry breaking vacua. The distance between $\langle \hat \Phi
\rangle |_{\rm lower}$ and $\langle \hat \Phi \rangle |_{\rm ISS}$ is
of order $O(\bar m^2/m_z)$ and is wide compared
to the height of the potential. Thus, we can estimate bounce action with
triangle approximation \cite{Duncan},
\begin{eqnarray}
S\sim {\left(\frac{\bar m }{ \bar \mu }\right)^4\left(\frac{\bar m }{ m_z}\right)^4}. \nonumber 
\end{eqnarray}
Even if we choose $m_z\sim \bar m$, which will be required below, the
Euclidean action can be made parametrically large by taking $\bar \mu \ll
\bar m$. Thus, the decay rate is parametrically small. 

One might think that we can find more efficient path through tree level
potential barrier. However at least it has to climb up $V_{\rm peak}\sim
O(\bar \mu^2 \bar m^2)$ that is very high, compared to the difference
between two supersymmetry breaking vacua, of order $O(\bar \mu^4)$. In
this case, we can use the thin wall approximation \cite{Coleman} to
estimate the bounce action and obtain $S \sim ( \bar m / \bar \mu)^8$.
Again, we can make it parametrically large when $\bar \mu \ll \bar m$.


\vspace{0.3cm}

{\it \large Supersymmetry preserving vacua}

\noindent So far we studied supersymmetry breaking vacua. In addition to
these, the model also has supersymmetric vacua. Here, we will show that
these supersymmetry preserving vacua can also be identified in the free
magnetic dual description. Following \cite{Intriligator:2006dd}, we look
for a supersymmetric vacuum where meson fields get large expectation
values. By the vacuum expectation value of $Y$ and $\hat{\Phi}$, dual
quarks $\chi,\tilde{\chi}$ and $\rho, \tilde{\rho}$ become massive and
can be integrated out. Also in the energy scale $E< hm_z$, $Z$ and
$\tilde{Z}$ should be integrated out. Thus, we are left with the
superpotential,
\begin{eqnarray}
W=-h \bar m^2Y -h \bar \mu^2 \hat{\Phi}+(N_F-N_C)\Lambda_{\rm eff}^3. \nonumber 
\end{eqnarray}
where the last term is generated by non-perturbative dynamics of a pure
SU$(N_F-N_C)$ gauge theory. The low energy scale $\Lambda_{\rm eff}$
after decoupling of dual quarks, is given by the matching conditions at
the two mass scales $hY$ and $h\hat{\Phi}$,
\begin{eqnarray}
\Lambda_{\rm eff}^{3}=\langle hY \rangle \langle h\hat{\Phi} \rangle^\frac{N_C }{ N_F-N_C}\Lambda_m^\frac{2N_F-3N_C }{ N_F-N_C}\ . \nonumber 
\end{eqnarray}
Note that $Z$ and $\tilde{Z}$ are singlets for the gauge group and do not
contribute to running of gauge coupling. With the non-perturbative
superpotential, $F$-term conditions for light field $Y$ and $\hat{\Phi}$
have solutions of the form,
\begin{eqnarray}
\langle h \hat{\Phi} \rangle=\bar m^\frac{2(N_F-N_C)}{N_C}\Lambda_m^\frac{3N_C-2N_F }{ N_F-N_C}
,\qquad \langle hY \rangle =\frac{\bar \mu^2 }{ \bar m^\frac{2(2N_C-N_F)}{ N_C}}\Lambda_m^\frac{3N_C-2N_F}{N_C}.\nonumber 
\end{eqnarray}
Since $\langle h\hat{\Phi} \rangle \gg \langle hY \rangle$ and the
difference of the vacuum expectation value $\hat{\Phi}$ between supersymmetric vacua and
supersymmetry breaking vacua is very large, compared to the height of
supersymmetry breaking vacua, we can estimate the Euclidean action for
the decay process by triangle approximation \cite{Duncan},
\begin{eqnarray}
S\sim \frac{\langle h \hat{\Phi}\rangle^4 }{ \bar \mu^4}\sim 
\left(
\frac{\bar m}{ \bar \mu}
\right)^4
\left(
\frac{\Lambda_m}{\bar m}
\right)^{4(3 N_C - 2 N_F)/N_C}\ .
\end{eqnarray}
The factor $3N_C - 2 N_F$ is always positive.
Therefore, with the mass hierarchy $\bar \mu \ll \bar m$ and $\bar m \ll
\Lambda_m$, we can make the action
arbitrarily large, and thus make the meta-stable vacua arbitrarily
long-lived. These conditions also allow us to ignore higher order
correction to the K\"ahler potential.

\subsection{Gaugino and scalar masses}

With the explicit $R$-symmetry breaking by $m_z$, direct mediation of
supersymmetry breaking happens. The standard model gauge group can be
embedded into either the SU($N_F - N_C$)$_F$ or the SU($N_C$)$_F$ flavor
symmetry which is remained unbroken at low energy.

The gaugino masses are, in this case, given by the same formula in
Eq.~(\ref{eq:gaugino}) with mass matrix ${\cal M}$:
\begin{eqnarray}
 {\cal M} = \left(
\begin{array}{cc}
 \hat \Phi & \bar m \\
 \bar m & m_z \\
\end{array}
\right)\ .
\end{eqnarray}
Therefore
\begin{eqnarray}
 m_\lambda = \frac{ g^2 \bar N }{(4 \pi)^2} \frac{h \bar \mu^2}{\bar
  m} \frac{ m_z }{ \bar m}
+ O\left( \frac{m_z^2}{\bar m^2} \right)\ ,
\end{eqnarray}
with $g^2$ the gauge coupling constant of the standard model gauge
interaction. The factor $\bar N$ is again $\bar N = N_C$ $(\bar N=N_F -
N_C)$ when we embed the standard model gauge group into SU($N_F - N_C$)
(SU($N_C$)).

Scalar masses are also obtained by two-loop diagrams. It is calculated
to be
\begin{eqnarray}
 m_i^2 = 2 \bar N C_2^i
\left(
\frac{g^2}{(4 \pi)^2}
\right)^2
\left( \frac{h \bar \mu^2}{\bar m} \right)^2
+ O\left( \frac{m_z^4}{\bar m^4} \right)\ .
\end{eqnarray}
$C_2^i$ is a quadratic Casimir factor for a field labeled $i$.
For having a similar size of gaugino and scalar masses, $m_z \sim \bar
m / \sqrt{ \bar N}$ is required. It is possible to have this relation as long as $m_z <
\bar m$ without destabilizing the meta-stable vacuum.

\subsection{Mass spectrum and the Landau pole problem}

We summarize the mass spectrum at the ISS vacuum here. 
The massless modes are the Goldstone boson, Im$[\tr \delta \hat \chi]$, and
the fermionic component of $\tr \delta \hat \chi$.
The pseudo-moduli $\hat \Phi$ and Re$[\tr \delta \hat \chi]$ have masses
which are similar size to the gauginos, i.e., $O(100~{\rm GeV})$. Other
component fields in the chiral multiplets $Y$, $Z$, $\tilde Z$, $\rho$,
$\tilde \rho$, $\chi$ and $\tilde \chi$ have masses of $O(h \bar m)$ or
eaten by the gauge/gaugino fields.

Discussion of the Landau pole depends on a way of embedding of the
standard model gauge group into flavor symmetries. We separately discuss
two cases. We find that it is possible to avoid a Landau pole if we
embed the standard model gauge group into the SU($N_F - N_C$)$_F$ flavor
symmetry and take the dynamical scale and the mass parameter $\bar m$ to
be large enough.
We also comment on an alternative possibility that the SU($N_C$) gauge
theory above the scale $m$ is a conformal field theory (CFT). This
possibility allows us to take the mass parameter $m$ and the
dynamical scale $\Lambda$ to be much lower than the unification scale
without the Landau pole problem.

\subsubsection{ Embedding SU(3) $\times$ SU(2) $\times$ U(1) into
   SU(${\mathbf{ N_F - N_C }}$)$_\mathbf{F}$ }

In this case, the pseudo-moduli $\hat \Phi$ is a singlet under the
standard model gauge group, and thus it does not contribute to the beta
function.

The beta function coefficients of the SU(3) gauge coupling is
\begin{eqnarray}
 b_3 (\mu_R < h \bar m ) = -3, \quad 
 b_3 ( h \bar m < \mu_R < \Lambda ) = - 3 + 2 N_F - N_C, \quad
 b_3 ( \mu_R > \Lambda) = -3 + N_C,
\end{eqnarray}
where $\mu_R$ is a renormalization scale.
Above the mass scale $m_X (\gg \Lambda)$, which is defined in
Eq.~(\ref{eq:higher-d}), the theory should be replaced by a
renormalizable theory, where it neccesarily contains additional fields.
Therefore, there are contributions from those fields above the scale
$m_X$. The size of the contributions depends on a specific ultra-violet
completion of the theory.

In order for the embedding to be possible, $N_F - N_C \geq 5$, and from the
condition $N_C < N_F < {3\over 2}N_C $, we obtain
\begin{eqnarray}
 2 N_F - N_C > 20, \quad N_C > 10\ .
\end{eqnarray}
There is a quite large contribution to the beta function. 
To avoid a Landau pole below the unification scale, $M_{\rm GUT} \sim
10^{16}$~GeV, the mass scales $h \bar m$ and $\Lambda$ should be high
enough. For example, $\Lambda \sim M_{\rm GUT}$ and $h \bar m \gtrsim
10^{13}$~GeV can avoid the Landau pole.

\vspace{0.3cm}
{\it Model in the conformal window} 

\noindent
Although it is not conclusive, the authors of
Ref.~\cite{Intriligator:2006dd} suggested that there is a meta-stable
supersymmetry breaking vacuum also when the numbers of colors and flavors
are the same. If it is the
case, there is an interesting possibility that we can go into the
conformal window, ${3\over 2}N_C \leq N_F < 3 N_C$.
If $N_F$ is in the conformal window, the gauge coupling of SU($N_C$)
flows into the conformal fixed point at some scale $\Lambda_*$. The
theory stays as a CFT until the mass term $m (Q_I \bar Q_I)$ becomes
important, and eventually at a lower scale $\Lambda \sim m$, the theory
exits from the CFT and becomes strongly coupled.
The effective theory below the scale $\Lambda \sim m$ is described by an
SU($N_C$) gauge theory with $N_C$ flavors with a mass term $\mu (Q_a
\bar Q_a)$. This is exactly the ISS model with $N_C$ flavors.
Once we assume the existence of the meta-stable supersymmetry breaking
vacuum, direct gauge mediation should happen as we discussed in the
previous section although we have lost the control of the perturbative
calculation. (See \cite{Izawa:2005yf} for a similar model.)
The beta function coefficient $b_3$ is in this case,
\begin{eqnarray}
 b_3 (\mu_R < \Lambda ) = - 3, \quad
 b_3 (\Lambda < \mu_R < \Lambda_* ) = - 3 + \frac{3 N_C^2}{N_F} + \Delta, \quad
 b_3 (\mu_R > \Lambda_* ) = - 3 + N_C + \Delta^\prime \ ,
\label{eq:beta-cft}
\end{eqnarray}
where we have included a contribution from anomalous dimensions of $Q$'s
in CFTs~\cite{Novikov:1982px,Shifman:1986zi,Seiberg:1994bz}. The factors
$\Delta$ and $\Delta^\prime$ are unspecified contributions from the
fields which generate the $m_z$ term.
With ${3\over 2}N_C \leq N_F < 3 N_C$ and $N_F - N_C \geq 5$, we find
\begin{eqnarray}
 N_C \geq 3, \quad
 N_F \geq 8, \quad
 {3 N_C^2 \over N_F} \geq {27\over 8}\ .
\end{eqnarray}
Therefore, the dynamical scale $\Lambda \sim m$ can be much lower than
the unification scale in this case. 
For example, if we take the ultra-violet completion to be simply adding
a pair of massive fields $\eta_{Ia}$ and $\tilde \eta_{aI}$ which couple to
$(Q_a \bar Q_I)$ and $(Q_I \bar Q_a)$, respectively, the additional
contributions are $\Delta = \Delta^\prime = N_C$. In this case, we can
take the dynamical scale $\Lambda \sim m$ to be as low as
$O(100-1000~{\rm TeV})$ without a Landau pole problem for $N_C = 3$ and
$N_F = 8$.

We implicitly took the scale $m_X$, where the $m_z$ term is generated,
to be $O(\Lambda)$ in Eq.~(\ref{eq:beta-cft}) because of the requirement
$m_z \sim \bar m$ for the sizes of the gaugino and scalar masses to be
similar. With $m_z \sim \Lambda^2 / m_X$ (see Eq.~(\ref{eq:higher-d}))
and $m \sim \Lambda$, we need to take $m_X \sim \Lambda$.
However, the actual scale at which new fields appear can be much higher
than $\Lambda$ or even $\Lambda_*$ when the anomalous dimensions of $Q$
and $\bar Q$ are large in the CFT.
For example, when $N_F \leq 2 N_C$, $(Q_I \bar Q_a) (Q_a \bar Q_I)$ is a
marginal or a relevant operator. In this case, it is not required to
have an ultra-violet completion of the theory up to $O(\Lambda_*)$ or
higher, i.e., $\Delta = 0$, while satisfying $m_z \sim \bar m$.
This can be understood by the running of the $1/m_X$ parameter in the
CFT:
\begin{eqnarray}
 \frac{1}{m_X (\mu_R)} = \frac{1}{ m_X (\Lambda) }
\left(
\mu_R \over \Lambda
\right)^{(2N_F - 6 N_C)/N_F}\ .
\end{eqnarray}
The unspecified contribution $\Delta^\prime$ is not important if
$\Lambda_*$ is high enough.

If $N_F - N_C > 5$, there are flavors with mass $m$ which are not
charged under the standard model gauge group. If we reduce the masses of
those fields to be slightly smaller than $m$, the low energy effective
theory below $\Lambda$ has more flavors and we can perform a reliable
perturbative calculation of the potential for pseudo-moduli.

It is interesting to note that this CFT model may be regarded as a dual
description of models with a warped extra-dimension in
Refs.~\cite{Gherghetta:2000qt,Gherghetta:2000kr,Goldberger:2002pc,Nomura:2004zs},
where supersymmetry is broken on an infrared brane, and standard model
gauge fields are living in the bulk of the extra-dimension.

\subsubsection{ Embedding SU(3) $\times$ SU(2) $\times$ U(1) into
   SU(${\mathbf{ N_C }}$)$_\mathbf{F}$ }

In this case, $b_3$ is given by
\begin{eqnarray}
 b_3 (\mu_R < h \bar m ) = -3 + N_C, \quad 
 b_3 ( h \bar m < \mu_R < \Lambda ) = - 3 + 2 N_F - N_C, \quad
 b_3 ( \mu_R > \Lambda) = -3 + N_C.
\end{eqnarray}
The condition for the embedding to be possible is $N_C \geq 5$. Therefore
\begin{eqnarray}
 2 N_F - N_C > 5, \quad N_C \geq 5\ .
\end{eqnarray}
With this constraint, there is always a Landau pole below the
unification scale. 
The situation does not improve even if we consider the possibility of
the CFT above the mass scale $m$.

\bigskip 
To summarize, by embedding the standard model gauge group in the SU($N_F
- N_C$)$_F$ subgroup of the flavor symmetry, we can couple the ISS model
to the standard model. The gaugino masses are generated at one-loop, and
the Landau pole problem can be avoided if the gauge coupling scale of the ISS
sector is sufficiently high or if the theory above the mass scale $m$ is a
CFT.

\section{Ultra-violet completions}

The perturbation to the ISS model we considered in the previous section
is non-renormalizable in the electric description.  
In this section we will show that the model can be regarded as a low
energy effective theory of a renormalizable gauge theory at high energy. 
Moreover, this renormalizable theory itself can be realized as a 
low energy effective theory on intersecting branes and on branes on 
a local Calabi-Yau manifold in string theory. 
In order to decouple Kaluza-Klein and string excitations from the 
gauge theory, the length scale of 
these brane configurations as well as the string length 
must be smaller than that of the gauge theory.
These brane configurations are so simple that it may be possible to 
incorporate them in the on-going effort to construct the minimal 
supersymmetric standard model from string theory compactifications. 

One way to generate the non-renormalizable
interaction (\ref{eq:higher-d}) is as follows. 
Consider an ${\cal N}=2$ quiver gauge theory with the gauge group 
U($N_1$) $\times$ U($N_2$) $\times$ U($N_3$) with
\begin{eqnarray}
  N_1=N_F-N_C,~N_2=N_C,~N_3=N_C,
\end{eqnarray}
and identify U($N_2$) with the gauge group U($N_C$) of the ISS
model.\footnote{In the previous sections, we consider the case when the
gauge group is SU($N_C$). When the gauge group is U($N_C$), the
``baryon'' symmetry is gauged and one of the pseudo-moduli ${\rm Tr}
\delta \hat \chi$ becomes massive at tree-level due to the additional
D-term condition. Otherwise, there is no major difference in properties
of meta-stable vacua.}  We assume that the scales $\Lambda_1, \Lambda_3$
for the other gauge group factors are so low that we can treat U($N_1$)
$\times$ U($N_3$) as a flavor group.  We then deform the theory by
turning on the superpotential $W_1(X_1)+W_2(X_2)+W_3(X_3)$ for the
adjoint fields $X_1, X_2, X_3$ in the ${\cal N}=2$ vector multiplets
given by
\begin{eqnarray}
W_1=\frac{M_X}{2}X_1^2+\alpha_1 X_1, \quad
W_2=-\frac{M_X}{2}X_2^2, \quad
W_3=\frac{M_X}{2}X_3^2+\alpha_3 X_3 .\nonumber
\end{eqnarray}
This breaks ${\cal N}=2$ supersymmetry into ${\cal N}=1$, 
and the total tree level superpotential 
of the deformed theory is
\begin{align}
W_{tree}=&-Q_{21}X_1Q_{12}+Q_{12}X_2Q_{21}-Q_{32}X_2Q_{23}+Q_{23}X_3Q_{32}\nonumber \\ 
& +\, W_1(X_1)+W_{2}(X_2)+W_3(X_3) \nonumber 
\end{align}
After integrating out massive fields $X_i$, the superpotential can be written as\begin{align}
&W_{tree} = {\tr} \, m_Q Q\bar{Q} +{\tr} \, K_1 Q\bar{Q} K_2 Q\bar{Q} \nonumber \\
 m_Q&={\rm diag}\left({\alpha_1/M_X},{\alpha_3/M_X}\right),\quad K_1={\rm diag}\left( 0, {1/M_X} \right), \quad K_2={\rm diag}\left( 1, 0 \right). \nonumber
\end{align}
This reproduces the interaction (\ref{eq:higher-d}) and the mass terms for 
$(Q_I, Q_a)$ if we set
\begin{eqnarray}
\frac{\alpha_1 \Lambda_2}{M_X}  = h \bar m^2 ,\qquad 
\frac{\alpha_3 \Lambda_2}{M_X}  = h \bar \mu^2,\qquad 
\frac{\Lambda_2^2}{M_X}=hm_z . \label{set}
\end{eqnarray}
Since we suppose $\Lambda_2 < M_X$, all the equations \eqref{set} can be
satisfied by appropriately choosing parameters $\alpha_{1,2}$ and $M_X$.

\bigskip

\subsection{Embedding in string theory}

In the perturbative string theory, the collective coordinates of D-branes
are open strings ending on them \cite{Polchinski}. Since the lightest
degrees of freedom of open strings include gauge fields, variety of 
gauge theories arise on intersecting branes in the low energy limit
where the string length becomes small and the coupling of D-branes 
to the bulk gravitational degrees of freedom becomes negligible 
\cite{HW,EGK,GK}. We will present an intersecting brane configuration
where the deformed ISS model is realized as a low energy effective theory. 
One should not be confused that our use of the intersecting brane model 
implies that the theory above the dynamical scale $\Lambda$ is replaced 
by string theory or a higher dimensional theory. The string length and
the compactification scale are much shorter than the gauge theory scale. 
It is one of the string miracles that quantum moduli spaces of low energy 
gauge theories are often realized as actual physical spaces such as brane 
configurations or Calabi-Yau geometry, allowing us to discuss deep infrared 
physics in the ultra-violet descriptions of the theories.
This phenomenon has been well-established for moduli spaces of
supersymmetric vacua, and it has just begun to be explored for
supersymmetry breaking vacua \cite{O2II,FU,IAS,ABFK,ABSV,Verlinde}. 
(For earlier works 
in this direction, see for example \cite{dBHOO,Vafa:2000,Kachru2002}.)
Here, we will find
that meta-stable supersymmetry breaking vacua of the deformed ISS model 
are realized as geometric configurations of branes. 
 
Consider Type IIA superstring theory in 
the flat 10-dimensional Minkowski spacetime with coordinates
$x^{0,\cdots,9}$.  Introduce four NS5 branes located at $x^{7,8,9}=0$
and at different points in the $x^6$ direction, and extended in the
$x^{0,\cdots,3}$ and $x^{4,5}$ directions.  Let us call these NS5 branes
from the left to right along the $x^6$ direction as NS5$_1$, NS5$_2$,
NS5$_3$, and NS5$_4$.  We then suspends $(N_F-N_C)$ D4 branes between
NS5$_1$ and NS5$_2$, $N_C$ D4 branes between NS5$_2$ and NS5$_3$, and
$N_C$ D4 branes between NS5$_3$ and NS5$_4$. The brane dynamics in the
common $x^{0,\cdots,3}$ directions is described by the ${\cal N}=2$
supersymmetric 
quiver gauge theory with the gauge group U($N_F-N_C$) $\times$ U($N_C$)
$\times$ U($N_C$). Note that the gauge coupling constant
$g_{{\rm YM}}^{(i)}$ $i=1,2,3$ for the three gauge group factors
are given at the string scale by
\begin{eqnarray} (g_{{\rm YM}}^{(i)})^2 = g_{\rm s} {\ell_{\rm s} \over L_i},
\end{eqnarray}
where $g_{\rm s}$ and $\ell_{\rm s}$ are string coupling constant and
string length, and $L_1, L_2, L_3$ are the lengths of the three types of 
D4 branes suspended between NS5 branes. The gauge couplings 
$g_{{\rm YM}}^{(i=1,2,3)}$ set the initial conditions for the 
renormalization group equation at ultra-violet. The ${\cal N}=2$
quiver gauge theory
is realized in the low energy limit where $g_s, \ell_{\rm s}, 
L_i \rightarrow 0$, keeping $g_{\rm YM}^{(i)}$ fixed. We choose
$L_2 \ll L_1, L_3$ so that the gauge coupling constants for
U($N_1$) $\times$ U($N_3$) are small. 

\begin{figure}[htbp]
\begin{center}
  \epsfxsize=11cm \epsfbox{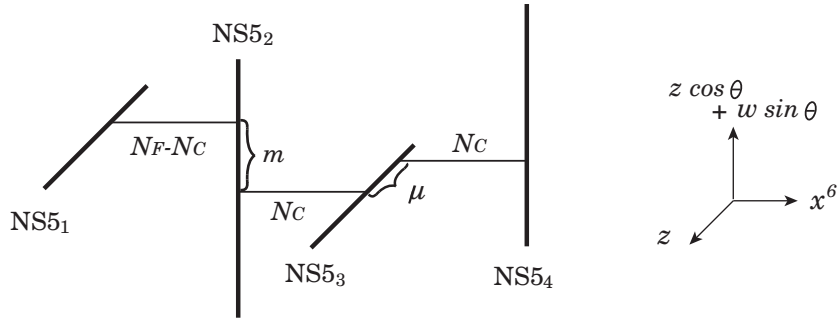}
\end{center}
 \vspace{-.5cm}
\caption{\sl The electric brane configuration.   \label{figelectric}}
\end{figure}

We can turn on the superpotentials $W_1+W_2+W_3$ by rotating NS5$_2$
and NS5$_4$ into the $x^{7,8}$ directions. More
precisely, we use the complex coordinates $z=x^4+ix^5$ and $w=x^7+ix^8$
and rotate the two NS5 branes on the $z-w$ plane so that they are
extended in the direction of $\cos \theta z + \sin\theta w$.  The
holomorphic rotation preserves ${\cal N}=1$ supersymmetry.  In the field
theory, this corresponds to turning on $W_1+W_2+W_3$ with $M_X =
\tan\theta$ \cite{barbon}.  We can also turn on the quark masses $m$ and
$\mu$ by moving NS5$_1$ and NS5$_4$ in the $w$ direction. The resulting
configuration is shown in Figure \ref{figelectric}.

We can also T-dualize the NS5 branes to turn the D4 branes suspended
between the NS5 branes into D branes wrapping compact cycles in a local
Calabi-Yau manifold \cite{CFIKV}. Realizations of meta-stable vacua 
on branes partially wrapping cycles in Calabi-Yau manifolds have been
discussed, for example, in \cite{OO,ABSV}.

The brane configuration shown in Figure 2 is similar to the one appeared
recently in \cite{ABFK}. However, there are some important
differences. In the model of \cite{ABFK}, the quark masses $m$ and $\mu$
in the electric description are set equal to zero. Moreover the strong
coupling scales of the three gauge group factors are chosen as
$\Lambda_1, \Lambda_2 \ll \Lambda_3$ in the model of \cite{ABFK},
whereas $\Lambda_1, \Lambda_3 \ll \Lambda_2$ in our model.  These
differences have led to different ways of supersymmetry breaking in
these models. Despite the differences, some of the results in
\cite{ABFK} may be useful for further studies of our model.

\subsection{Meta-stable supersymmetry breaking vacua on the brane
configuration}

In \cite{O2II,FU}, the ISS model and its magnetic dual were studied by
realizing them on intersecting branes, and brane configurations for the
supersymmetry breaking vacua were identified. The brane
configurations provide a geometric way to understand the vacuum structure of
the model.\footnote{See \cite{IAS} on
issues that arise when one turns on finite string coupling in  
these brane configurations. These issues are not relevant to our 
discussion below since we mostly deal with tree-level
properties of Type IIA superstring theory.}
Recently it was used, for example, to study solitonic states
on the meta-stable vacuum in the ISS model \cite{Eto:2006yv}. Here, we
will present brane configurations that correspond to the meta-stable
vacua in the deformed ISS model.

To identify the meta-stable vacua, we need to go to the magnetic
description, which is realized on branes by exchanging NS5$_2$ and
NS5$_3$. Since we assume $L_2 \ll L_1, L_3$, it is reasonable to 
expect that the first duality transformation involves only these
two NS5 branes. To avoid confusion, let us call the resulting NS5 branes as
NS5$_1$, NS5$_2'$, NS5$_3'$, and NS5$_4$ from the left to right in the
$x^6$ direction. Note that NS5$_1$ and NS5$_2'$ are parallel to each
other, and so are NS5$_3'$ and NS5$_4$. There are $(N_F-N_C)$ D4 branes
between NS5$_1$ and NS5$_3'$, $N_C$ anti-D4 branes between NS5$_2'$ and
NS5$_3'$, and $N_C$ D4 branes between NS5$_2'$ and NS5$_4$.

\begin{figure}[htbp]
\begin{center}
  \epsfxsize=11cm \epsfbox{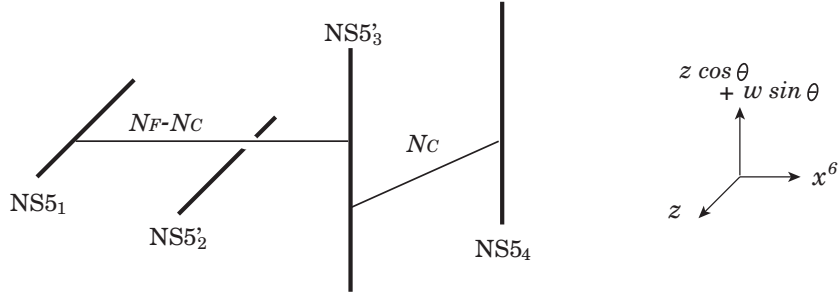}
\end{center}
 \vspace{-.5cm}
\caption{\sl The brane configuration for the ISS vacuum
in the deformed ISS model. \label{figISS}
}
\end{figure}

The ISS vacuum is obtained by bending the $N_C$ D4 branes between NS5$_2'$ 
and NS5$_4$ toward NS5$_3'$, disconnect each of them at NS5$_3'$,
and annihilate their segments between NS5$_2'$ 
and NS5$_3'$ with the $N_C$ anti-D4 branes by the tachyon condensation. The 
resulting brane configuration is shown in Figure \ref{figISS}. Note that this configuration
breaks supersymmetry since the D4 branes between NS5$_1$ and NS5$_3'$ and the 
D4 branes between NS5$_3'$ and NS5$_4$ are in angles. Since their end-point
separation is of the order of $|m|$ whereas the supersymmetry breaking is
of the order of their relative angles $\sim |\mu|$,  
an open string stretched between these D4 branes does not contain a tachyon
mode provided $|m| \gg |\mu|$. 
Since NS5$_3'$ and NS5$_4$ are parallel to each other, the $N_C$ D4 branes
between them can move along them. This freedom corresponds to pseudo-moduli
$\hat \Phi$. These D4 branes are stabilized by a potential induced by 
closed string exchange between them and the D4 branes between NS5$_1$ and
NS5$_2'$, which is the closed string dual of the Coleman-Weinberg potential.

We can also identify the other meta-stable vacua of the deformed ISS model. 
Let us take $n$ of the $N_C$ D4 branes between NS5$_3'$ and NS5$_4$ and
move them toward the $(N_F-N_C)$ D4 branes between NS5$_1$ and NS5$_3'$.
Doing this costs energy since these D4 branes have to climb up the 
Coleman-Weinberg potential. Eventually, as they approach the D4 branes 
between NS5$_1$ and NS5$_3'$, open strings between the two kinds of D4 branes
start developing tachyonic modes. The tachyon condensation then reconnects
$n$ pairs of D4 branes, leading to the brane configuration as shown in Figure 4. 
This process lowers the vacuum energy since
the length of the single D4 brane between NS5$_1$ and NS5$_4$ 
is shorter than the sum of the two D4 branes before the reconnection.

\begin{figure}[htbp]
\begin{center}
  \epsfxsize=11cm \epsfbox{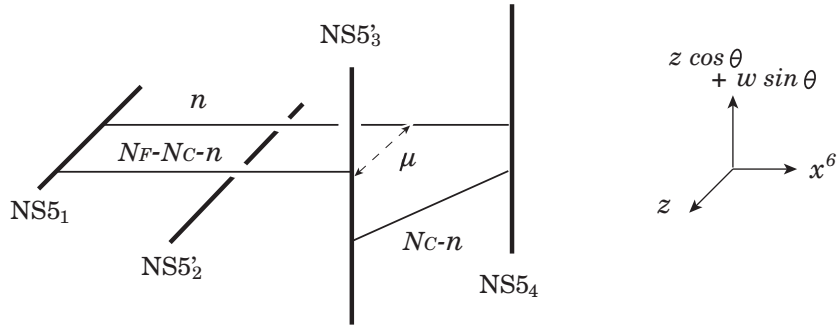}
\end{center}
 \vspace{-.5cm}
\caption{\sl The brane configuration for the meta-stable vacua with
lower energies.  \label{figLower}}
\end{figure}

One can show that these brane configurations
reproduce various features of the corresponding meta-stable vacua, such
as their vacuum energies, expectation values of various fields (such
as $\rho\tilde\rho$, $Y$, and $\hat\Phi$), and their decay processes.
This can be done by a straightforward application of the brane
configuration analysis in \cite{O2II,FU,IAS}, and we leave it as
an exercise for the readers. 

\section{Meta-stability at finite temperature?}

It has been shown that the meta-stable supersymmetry breaking vacuum in
the ISS model is favored in the thermal history of the
universe~\cite{reheatingone,reheatingtwo,reheatingthree}. The essential
observation is that there are more light degrees of freedom in the
supersymmetry breaking vacuum compared to the supersymmetric one. Finite
temperature effects make the meta-stable vacuum more attractive in this
circumstance.

In the deformed ISS model we discussed in this paper, there are many
other meta-stable vacua. However, interestingly, the desired vacuum (the
ISS vacuum) possesses the largest symmetry group among those vacua. In
other vacua, number of degrees of freedom of the pseudo-moduli is
reduced because some components of $\hat \Phi$ have masses at tree
level.
Therefore, the desired vacuum is the most attractive in the thermal
history of the universe.

\section*{Acknowledgments}

We would like to thank K.~Intriligator, H.~Murayama, E.~Silverstein,
T.~Watari, and T.~Yanagida for discussions. RK
thanks the hospitality of the high energy theory group at Rutgers
University.  The work of RK was supported by the U.S. Department of
Energy under contract number DE-AC02-76SF00515. The work of HO and YO is
supported in part by the U.S. Department of Energy under contract number
DE-FG03-92-ER40701.  HO is also supported in part by the U.S. National
Science Foundation under contract number OISE-0403366.  YO is also
supported in part by the JSPS Fellowship for Research Abroad.


\end{document}